# Numerical simulation of two-dimensional detonation propagation in partially pre‑vaporized *n*-heptane sprays

Majie Zhao, Huangwei Zhang*
Department of Mechanical Engineering, National University of Singapore, 9 Engineering Drive 1, Singapore 117576, Republic of Singapore
*Corresponding author email: huangwei.zhang@nus.edu.sg

**Abstract**
In this paper, two-dimensional detonation propagation in partially pre-vaporized *n*-heptane sprays is studied by using Eulerian–Lagrangian methods. The effects of droplet pre-evaporation on the detonation propagation are investigated. The general features and detailed structures of two-phase detonations are well captured with the present numerical methods. The results show that the detonation propagation speed and detonation structures are significantly affected by the pre-evaporated gas equivalence ratio. The numerical soot foils are used to characterize the influence of pre-evaporated gas equivalence ratio on the detonation propagation. Regular detonation cellular structures are observed for large pre-evaporated gas equivalence ratios, but when decreasing the pre-evaporated gas equivalence ratio, the detonation cellular structures become much more unstable and the average cell width also increases. It is also found that the pre-evaporated gas equivalence ratio has little effects on the volume averaged heat release when the detonation propagates stably. Moreover, the results also suggest that the detonation can propagate in the two-phase n-heptane/air mixture without pre-evaporation, but the detonation would be first quenched and then re-ignited when the pre-evaporated gas equivalence ratio is small or equal to zero.

**Introduction**
Detonation is a shock-induced combustion wave with high-energy release rate and high pressure. Due to its promising application in propulsion, such as the pulse detonation engine (PDE) and the rotating detonation engine (RDE), the detonation structures and propagation characteristics have been investigated by many researchers. In previous studies, most work are mainly focused on the detonation propagation in gaseous fuel/oxidizer mixtures. However, realistic hydrocarbon liquid fuels should be used in future practical applications, due to their higher energy density and easier storage. Therefore, it is necessary to study the behaviours of detonation propagation in liquid hydrocarbon fuel/oxidizer mixtures.

Recently, with the improvement of computational conditions and experimental methods, as well as the increasing interests in the new concept detonation propulsion technology, the two-phase detonation combustion has received more and more attention and investigation. The detonation propagation regimes in liquid sprays were experimentally and numerically studied by Veyssiere et al. [1], and the results suggested that the detonation can propagate in volatile fuel sprays with sufficiently small droplet size. The detonation pressure, wave speed, and cell width in JP-10 mixtures were experimentally measured by Austin and Shepherd [2]. The Eulerian–Lagrangian methods were used by Schwer et al. [3][4] to study the detonation propagation in liquid JP-10/Oxygen mixtures, and some of the differences between gaseous fuel and liquid fuel detonations were addressed. The numerical results suggested that the cellular structures for multidimensional liquid fuel sprays were affected by the fuel spray droplet size due to the evaporation time variation, and the detonation propagation speed was decreased in the presence of droplets when compared to the case with fully mixed gaseous. The oblique detonations in two-phase kerosene–air mixtures were numerically studied by Ren et al. [5], and the results showed that the fuel-lean mixtures would be more sensitive to the droplets evaporation cooling effects, while the heat release effects predominate on the fuel-rich side. The detonation initiation and propagation in rotating detonation engine under the conditions of kerosene liquid mist and air injection were studied by Kindracki [6], and it was found that the measured rotating detonation propagation speed was lower than the theoretical C-J value by 20-25%. The continuous spin detonation of a heterogeneous kerosene–air mixture with addition of hydrogen was experimentally studied by Bykovskii et al. [7]. Our recent work [8][9] also suggested that the initial droplet diameter would have a significant impact on the rotating detonation propagation speed and detonation combustion efficiency. However, the instabilities and characteristics of detonation cellular structures in two-phase sprays, similar to that exhibited in gaseous mixtures, are still not well established.

In this work, we aim to investigate the influences of liquid *n*-heptane properties on the two dimensional detonation propagations. The effects of pre-vaporized gas equivalence ratio on the detonation propagation speed, detonation cellular structures, and the detonation instabilities will be discussed in detail. Here the initial droplet diameter of 10 μm is considered. The manuscript is organized as below. In Section 2 the computational method and the physical model are introduced. Results are presented in Section 3 and conclusions are made in Section 4.

**Governing equation and numerical method**

In this work, the Eulerian–Lagrangian method is used to investigate two-phase detonative combustion. For the gas phase, the governing equations of continuity, momentum, energy and species mass fraction, together with the ideal gas equation of state, are solved [10]. The liquid phase is modeled as a spray of spherical droplets tracked by Lagrangian method [11]. The inter-droplet interactions are neglected since dilute sprays (volume fraction < 0.001 [12]) are considered, and two-way coupling between gas and liquid phases are considered, in terms of mass, momentum, energy and species exchanges. More detailed information about the governing equations of both the gas and liquid phase can be found in our recent work [13].

Both the gas and liquid phase equations are solved by a multi-component, two-phase, and reactive solver, *RYrhoCentralFoam* [14], with two-way interphase coupling in terms of mass, momentum, energy and species. For the gas phase, second-order implicit backward method is employed for temporal discretization and the time step is about $1\times10^{-9}$s. The KNP (i.e. Kurganov, Noelle and Petrova [15]) scheme with van Leer limiter is used for MUSCL-type reconstructions of the convective fluxes in momentum equation. The second-order central differencing scheme is applied for the diffusion terms. It has been validated and successfully used for gaseous supersonic flows and detonative combustion problems [16–19]. For the liquid phase, they are solved using first-order implicit Euler method, which is sufficiently accurate with the time step used in this work (about $10^{-9}$s). Detailed validations and verifications of the solver and sub-models are performed in our recent work [13]. Satisfactory accuracies are achieved in predicting shock wave, detonation propagation speed, and detonation cell size. More information about the numerical schemes and solution strategies can be found in Refs. [10,13,20].

The chemical source terms are integrated with an Euler implicit method. Its accuracy can be confirmed through comparing with other advanced ODE integrators [13]. Two-step reaction of 6 species (*n*-$C_7H_{16}$, $O_2$, CO, $CO_2$, $H_2O$ and $N_2$) [21] is used for *n*-heptane detonation.

A rectangular computational domain (as shown in Fig. 1) is used. The length (x-direction) and height (y-direction) of the whole domain are 1000 mm and 40 mm, respectively. The initial pressure and temperature of *n*-heptane/air mixtures are 0.5 atm and 300 K, respectively. Hot spots with high temperature (2000 K) and pressure (50 atm) are used to ignite the detonation wave. In order to study the detonation propagation in partially pre-vaporized *n*-heptane sprays, the *n*-heptane droplets are uniformly distributed in the rectangular region with $L_x$ = 200 mm and $L_y$ = 40 mm at the right end of the computational domain (see Fig. 1). Here the long distance (800 mm) before the detonation propagating into the fuel sprays is used to minimize the detonation initiation effects. In this work, the effects of liquid droplet pre-evaporation will be examined in detail. The equivalence ratio in the two-phase section (see Fig. 1) is divided into two parts, i.e., the equivalence ratio of pre-evaporated n-heptane $\phi_g$ and liquid droplet equivalence ratio $\phi_l$. The $\phi_g$ of 0.0, 0.2, 0.4 and 0.6 will be considered, respectively. The $\phi_l$ is defined as the mass ratio of the droplets to the oxidizer normalized by the mass ratio of *n*-heptane vapor to air under stoichiometric condition, and ranges from 0.0 to 1.0. The initial droplet diameter considered in the present work is 10 µm. The initial temperature and density of the droplet are 300 K and 680 kg/m$^3$, respectively.

The domain in Fig. 1 is discretized with 4,000,000 Cartesian cells, and the cell size in the x-direction and y-direction are uniform at 0.1 mm. The mesh resolution is larger than the considered droplet diameters, which can ensure that the gas phase quantities near the droplet surfaces (critical for estimating the two-phase coupling, e.g. evaporation) can be well approximated using the interpolated ones at the location of the sub-grid droplet [22]. The left and right boundaries are assumed to be non-reflective, and free-slip wall at the upper and lower sides are enforced. It should be noted that the current main work is mainly focused on the detonation propagation in fuel sprays. Therefore, the boundary layer effect is not considered due to the complexity and computational cost of such physical problems with boundary layer.

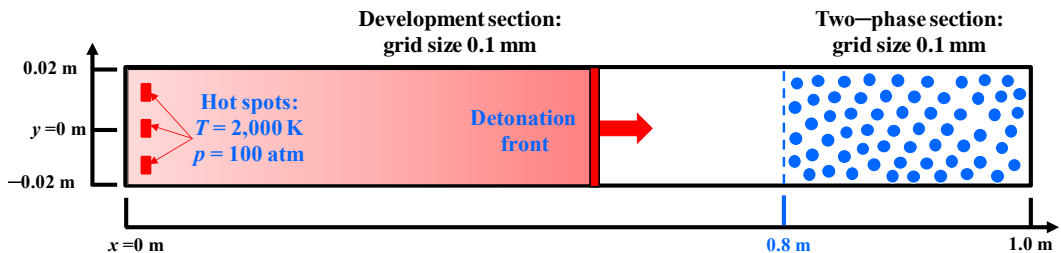

**Figure 1.** Computational domain of two-dimensional detonation propagation.

**Table 1** - Comparison with theoretical value.

| Equivalence ratio | Cellular size (mm) | | Propagation speed (m/s) | |
|---|---|---|---|---|
| | Present calculation | Theoretical value* | Present calculation | Theoretical value* |
| 0.6 | 5 ~ 26 | 31.2 | 1657 ± 323 | 1,603 |
| 1.0 | 4 ~ 10 | 7.4 | 1894 ± 251 | 1,802 |

*Calculation by using the SD Toolbox [23]

**Results and Discussion**

In order to validate the numerical methods and mesh resolution, the numerical simulations of detonation in pure gas *n*-heptane/air mixtures are performed before simulating the two-phase detonation propagation, and the results are compared with the theoretical value in terms of detonation propagation speed and cellular size. Here the equivalence ratio of 0.6 and 1.0 are considered. The comparison between the present numerical results and the theoretical values are summarized in Table 1. The results suggest that the present numerical methods and mesh size can be used to study the detonation propagation in n-heptane/air mixtures.

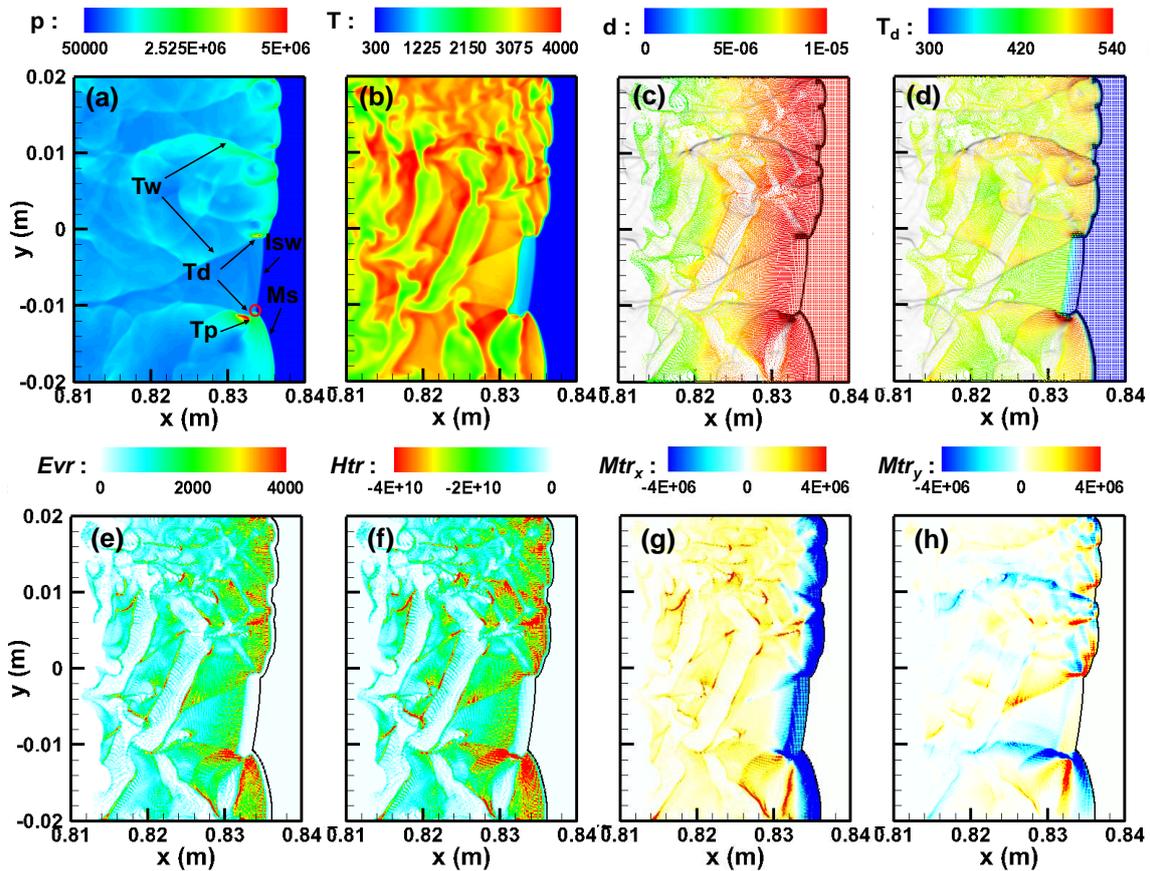

**Figure 2.** Distributions of (a) gas pressure (Pa), (b) gas temperature (K), (c) lagrangian *n*-heptane droplets colored by the (c) droplet diameter and (d) temperature, (e) evaporation rate $Evr$ (kg/m3/s), (f) heat transfer rate $Htr$ (J/m3/s), (g) momentum transfer rate in the x-direction $Mtr_x$ (N/m3) and (h) momentum transfer rate in the y-direction $Mtr_y$ (N/m3) in two-phase detonations. $\phi_g$ = 0.6, $\phi_l$ = 0.4 and d0 = 10 µm. Tw: transverse wave, Td: transverse detonation, Tp: triple point, Ms: Mach stem and Isw: incident shock wave. Black line in (e-f): leading shock front.

Figure 2 shows the distributions of gas pressure, gas temperature, droplet diameter, droplet temperature, evaporation rate, heat transfer rate, momentum transfer rate in the *x*- and *y*-directions, respectively, in two-phase detonations for the case with $\phi_g$ = 0.6, $\phi_l$ = 0.4 and d0 = 10 µm. It is shown that the basic detonation structures such as the Mach stem (Ms), incident shock wave (Isw), transverse wave (Tw) and triple point (Tp) are captured in the two-phase detonations. The pressure and temperature of n-heptane/air mixtures are increased behind the incident shock wave, and therefore, strong transverse detonations (Td) are formed. Stripe structures of gas temperature (see Fig 2b) are also observed behind the detonation front, which may be due to the interactions between the Mach stem, incident shock wave and the fuel droplets. The effects of the basic detonation structures on the fuel droplets can be observed with the distributions of droplets diameters and temperature (see Fig. 2c and 2d). The upward or downward movements of transverse waves and transverse detonations with large pressure gradient lead to the irregular distributions of the droplets. The fuel droplets would get heated behind the Mach stem, transverse waves and transverse detonations (see Fig. 2d), and therefore, the droplet diameter is slightly increased (see Fig. 2c) due to thermal expansion effect. However, compared to the droplet diameter and temperature in front of the detonation front, almost no changes are observed for the droplets behind the induced shock wave. As shown in Fig. 2(e) and 2(f), the droplet evaporation rate is significantly increased due to the high gas temperature behind the Mach stem, transverse shock wave and transverse detonation, where the heat transfer rate from the gas phase to the liquid droplet is also enhanced. For the momentum transfer rate in the x-directions (see Fig. 2g), it can be found that the droplets would be accelerated within a distance of about 2 mm behind the leading shock front, and then the droplets would be deaccelerated in a long distance of about 20 mm. Moreover, the droplets would be also accelerated upward or downward movement by the transverse waves (see Fig. 2h), which leads to the irregular distributions of the droplets (see Fig. 2c).

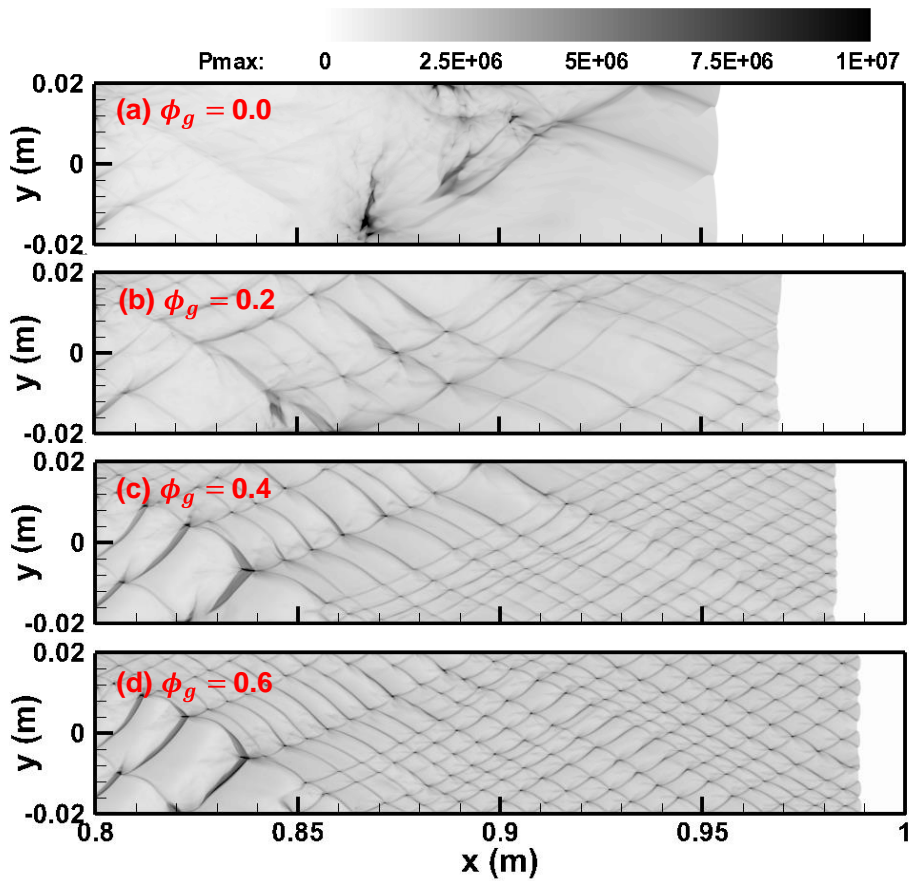

**Figure 3.** Evolution of maximum pressure with different initial liquid equivalence ratios at $t$ = 0.00055 s. The initial droplet diameter is 10 μm and $\phi_g + \phi_l = 1.0$.

Figure 3 shows the evolution of the detonation cellular structures in the two-phase section with different pre-evaporated gas equivalence ratios. Here the cases with initial droplet diameter of 10 μm and total equivalence ratio of $\phi_g + \phi_l = 1.0$ are considered. Note that the results shown in Fig. 6 are extracted at the same instant, i.e., $t$ = 0.00055 s. The location of the leading maximum pressure would represent the detonation propagation speed since they start at the same instant and with the same initial gas fields. The results with the evolution of the detonation cellular structures suggest that the detonation can propagate in the two-phase section without pre-evaporation (see Fig. 3a). It can also be found that the detonation would be first quenched and then re-ignited for the case without pre-evaporation. Nevertheless, with the increased initial pre-evaporated gas equivalence ratio, continuous detonation wave propagation is observed for the cases with some pre-evaporation, e.g., $\phi_g$ = 0.4 and 0.6. In addition, the cell width of the cellular structures decreases with the increased initial pre-evaporated gas equivalence ratio, and the detonation propagation will become much more stable. Therefore, the detonation propagation speed increases with the increased pre-evaporated gas equivalence ratio due to the stable detonation propagation.

As shown in Fig. 3(a), the detonation cellular structures show obvious instabilities, which suggests that the detonative combustion is also highly unstable. Here the volume averaged heat release rate is used to study the detonation instabilities when propagating. The volume averaged heat release rate is defined as the heat release rate in the whole two-phase computational domain, i.e., $\overline{Hrr} = \int_V Hrr dV$, where $Hrr$ is the local heat release rate and $V$ is the volume of the two-phase computational domain (see Fig. 1). Fig. 4 shows the time evolution of volume averaged heat release rate $\overline{Hrr}$ with different initial pre-evaporated gas equivalence ratios for cases with $d_0$ = 10 μm and $\phi_g + \phi_l$ = 1.0. When the detonation propagates into the two-phase section, the quenched detonation wave with low volume averaged heat release can be clearly observed for the cases with pre-evaporated gas equivalence ratio relatively small or even zero. However, with the increased pre-evaporated gas equivalence ratio, the quenching phenomenon of the detonation wave disappears, e.g., for the cases with $\phi_g$ = 0.4 and 0.6. Moreover, the time averaged heat release rate of re-ignited detonation wave is about 1.5×10⁵ J/s, which is very close to that with pre-evaporation of 0.6. This suggests that the effect of droplet pre-evaporation on the heat release rate is small when the detonation propagates stably.

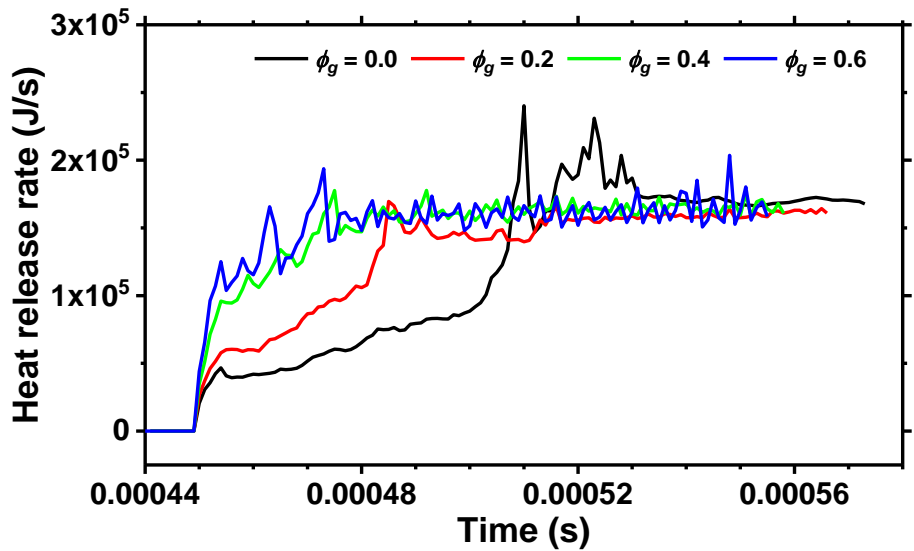

**Figure 4.** Time evolution of volume averaged heat release rate (J/s) with different initial droplet diameters for cases with the initial droplet diameter is 10 μm and $\phi_g + \phi_l$ = 1.0.

**Conclusions**

The Eulerian–Lagrangian methods with two-way gas−liquid coupling are used to study the two-dimensional detonation propagation in partially pre-vaporized *n*-heptane sprays. The effect of liquid droplet pre-evaporation on the detonation propagation is investigated. The numerical methods and mesh resolution are carefully validated by comparing the detonation propagation speed and cellular width with the theoretical value. The general features and detailed structures of two-phase detonations are well captured by the present numerical methods. It is found that the distributions of *n*-heptane droplets are also affected by the detonation wave structures. The droplets are first accelerated and then decelerated behind the detonation front, and the momentum exchange rate decreases with the increased initial droplet diameter.

The results of numerical soot foils suggest that the detonation propagation speed and cellular structures are significantly affected by the pre-evaporated gas equivalence ratio. Regular detonation cellular structures are observed for large pre-evaporated gas equivalence ratios, but when decreasing the pre-evaporated gas equivalence ratio, the detonation cellular structures become much more unstable and the average cell width also increases. Moreover, the results of the cases without droplet pre-evaporation suggest that the detonation can propagate in the two-phase n-heptane/air mixture without pre-evaporation, but the detonation would be first quenched and then re-ignited when the pre-evaporated gas equivalence ratio is small or equal to zero. It is also found that the pre-evaporated gas equivalence ratio has little effects on the volume averaged heat release when the detonation propagates stably.

**Acknowledgments**
The simulations used the ASPIRE 1 Cluster from National Supercomputing Centre, Singapore (NSCC) (https://www.nscc.sg/). This work is supported by MOE Tier 1 research grant (R-265-000-653-114).